\documentclass[review,12pt]{elsarticle}
\journal{ Physica A}
\usepackage[utf8]{inputenc}
\usepackage[T1]{fontenc}
\usepackage[english]{babel}

\usepackage{amsmath,amssymb}

\usepackage{graphicx}
\usepackage{dcolumn}
\usepackage{bm}
\usepackage{xcolor}
\usepackage[normalem]{ulem}

\usepackage[bookmarksnumbered,raiselinks,breaklinks]{hyperref}
\hypersetup{
colorlinks,
citecolor=blue,
filecolor=blue,
linkcolor=blue,
urlcolor=black
}

\usepackage[capitalise,nameinlink]{cleveref}

\usepackage{tcolorbox}
\tcbuselibrary{listings,breakable}

\definecolor{Myorange}{cmyk}{0,0.42,1,0}

\usepackage[normalem]{ulem}  

\begin{document}
\begin{frontmatter}
\title{Low-dimensional representation of brain networks for seizure risk forecasting}
\author[inst1]{Steven Rico-Aparicio}
\author[inst2]{Martin Guillemaud}
\author[inst2]{Alice Longhena}
\author[inst3,inst4]{Vincent Navarro}
\author[inst3,inst4]{Louis Cousyn}
\author[inst5]{Mario Chavez\corref{cor1}}
\cortext[cor1]{Corresponding author}
\ead{neurodynamicslab@gmail.com}
\address[inst1]{Physics Department, Industrial University of Santander, Colombia}
\address[inst2]{Paris Brain Institute (ICM), Inria Paris, INSERM-U1127, CNRS-UMR7225, Sorbonne University, Pitié-Salpêtrière Hospital, Paris, France}
\address[inst3]{Department of Neurology, Epilepsy Unit, Pitié-Salpêtrière University Hospital, Paris, France}
\address[inst4]{Paris Brain Institute, Sorbonne Université – CNRS UMR 7225 – Inserm U1127, Paris, France}
\address[inst5]{CNRS, Pitié-Salpêtrière Hospital, Paris, France}

\begin{abstract}
Identifying preictal states—periods during which seizures are more likely to occur—remains a central challenge in clinical computational neuroscience. In this study, we introduce a novel framework that embeds functional brain connectivity networks, derived from intracranial EEG (iEEG) recordings, into a low-dimensional Euclidean space. This compact representation captures essential topological features of brain dynamics and facilitates the detection of subtle connectivity changes preceding seizures. Using standard machine learning techniques, we define a dimensionless biomarker, $\mathcal{B}$, that discriminates between interictal (seizure-free) and preictal (within 24 hours of seizure) network states. Our method focuses on connectivity patterns among a subset of informative iEEG electrodes, identified through permutation-based testing, enabling robust classification of brain states across time. We validate our approach using a leave-one-out cross-validation scheme and a pseudo-prospective forecasting strategy, assessing performance with metrics such as F1-score and balanced accuracy. Results show that low-dimensional Euclidean embeddings of iEEG connectivity yield interpretable and predictive markers of preictal activity, offering promising implications for real-time seizure forecasting and individualized therapeutic interventions.
\end{abstract}

\end{frontmatter}

\section{Introduction}
The characterization of connectivity patterns in complex systems, from brain networks to large-scale infrastructures such as the Internet and social networks, exhibit intricate topologies that require sophisticated analytical frameworks capable of capturing their underlying structure and dynamics~\cite{boccaletti2006complex}. In the context of neuroscience, the study of intracranial electroencephalographic (iEEG) connectivity provides a crucial window into the functional organization of the brain, where network-based approaches have been employed to investigate seizure-related alterations in connectivity~\cite{van2014functional, cousyn2023daily}.

Various computational techniques, including coherence, mutual information, phase synchronization, and transfer entropy, have been applied to iEEG data to quantify neuronal interactions and detect patterns associated with epileptic activity~\cite{duncan2013intracranial}. A key aspect of seizure prediction research relies on comparing interictal (seizure-free) and preictal (seizure within the next 24h) states to identify early-warning signatures that may precede ictal events \cite{andrzejak2003testing, kuhlmann2018seizure}. However, one of the major challenges in this field is selecting an appropriate mathematical framework to represent and analyze brain connectivity networks.

Current approaches to predicting epileptic seizures largely rely on continuous EEG monitoring, using either scalp or intracranial recordings. This strategy demands substantial effort in both data acquisition and analysis. These models focus on detecting seizure-specific preictal patterns but are biased by changes related to shifts in vigilance states. As a result, they require either real-time identification of these states or the incorporation of diverse interictal reference periods.

Recent studies have debated whether Euclidean or non-Euclidean (e.g. elliptic, Gaussian, hyperbolic) embeddings provide a more suitable representation for functional brain networks \cite{allard2020navigable, boguna2021network}. While some non-Euclidean (hyperbolic) embeddings have been assumed to be superior in capturing non-trivial (hierarchical or multi-scale) connectivity structures~\cite{allard2020navigable}, their optimization is prone to numerical difficulties~\cite{inproceedingsGabriel}. Moreover, Euclidean embeddings offer a fundamental advantage: they provide an intuitive geometric representation with a well-structured distance metric in a space $\mathbb{R}^q$, facilitating a more interpretable analysis of embedded data (including the calculation of distances, barycenters and probability distributions).

The use of latent geometric representations of anatomical brain networks allowed to identify and quantify the impact of epilepsy surgery on brain regions~\cite{Longhena2024_chaos, Guillemaud2024_surgery}. In a recent study, we have explored hyperbolic mapping of brain networks to identify discriminative features relevant to seizure prediction~\cite{guillemaud2024hyperbolic}. These findings highlight the necessity of integrating network-based approaches to track seizure-related alterations and optimize therapeutic strategies.

Dimensionality reduction techniques play a pivotal role in network analysis, offering a statistical learning approach to condensing high-dimensional connectivity data while preserving essential structural and statistical information. This methodology applies to different brain network analyses, which typically follow a structured pipeline \cite{richiardi2013machine}: i) feature extraction: functional connectivity matrices are constructed using synchronization metrics such as the Phase Locking Value (PLV), mutual information, linear or non-linear correlation,\ldots; ii) graph reconstruction: connectivity matrices are transformed into network structures, where nodes represent EEG channels and edges reflect their functional relationships; iii) dimensionality reduction: high-dimensional network data is projected onto a lower-dimensional representation that conveys, with minimal distortion, both local and global geometric information of the original connectivity graph; and iv) classification and prediction: machine learning models are trained to differentiate between different cognitive or neurological states.

This study contributes to the field of computational neurosciences and epilepsy research by presenting a geometric framework that enables the iEEG  data representation and classification. This approach strengthens the analytical toolkits for studying epileptic dynamics by enabling the forecasting of state-dependent features and network nodes with high discriminatory value. In contrast to continuous monitoring methods, our strategy aims to assess whether an individual’s iEEG connectivity patterns, recorded during short resting states with controlled vigilance, show  differences between interictal (no seizure) and preictal (seizure within the next 24h) periods. We then explore the potential of these patterns to inform daily seizure risk predictions through the use of calibrated forecasting models.

The structure of this article is as follows: we first describe the dataset and preprocessing steps, then detail the embedding and alignment methods, followed by the biomarker formulation and classification strategy. We conclude with a presentation of the classification and forecasting results, and a discussion of their implications in the broader context of network neuroscience and clinical applications.

\section{Data acquisition}
The dataset consists of daily 10-minute resting-state intracranial EEG (iEEG) recordings from 10 patients (mean age: 30.7 years, mean duration of daily recordings: 11 days) with drug-resistant focal epilepsy, collected between January 2019 and July 2021 at the Epilepsy Unit of Pitié-Salpêtrière University Hospital (Paris, France). The study adhered to the Helsinki Declaration and was approved by an institutional review board (projects C11-16 and C19-55, sponsored by the National Institute of Health and Medical Research). 

Each 10-min resting-state period was labeled as ``inter-ictal'' if no seizure occurred in the next 24h ($N_1 = 69$), or ``preictal'' if at least one electroclinical seizure occurred in the next 24h ($N_2 = 38$). The implanted regions and the number of electrodes varied across patients, ranging from 10 to 62 electrodes (median = 20). For the purpose of the present study, only electrodes localized within the seizure onset zone (SOZ), as identified by the clinical team based on ictal activity, were retained for subsequent analyses. Details on the demographic information of each patient as age and number of recordings per day are available in the original study from this cohort~\cite{cousyn2023daily}.

\section{Brain networks construction}
Phase-locking value (PLV) was used to construct connectivity matrices from the iEEG data, which were computed between pairs of EEG signals during 20-second non-overlapping epochs, yielding 30 connectivity matrices per patient day. Synchrony values for the PLV were calculated for typical EEG frequency bands: delta ($\delta$, 1--4 Hz), theta ($\theta$, 4--8 Hz), alpha ($\alpha$, 8--13 Hz), beta ($\beta$, 13--30 Hz), low gamma (low $\gamma$, 30--49 Hz), and high gamma (high $\gamma$, 51--90 Hz). Only contacts located in the gray matter were considered, and a bipolar montage was applied between adjacent contacts to reduce any spurious synchrony produced by the volume conduction.

\subsection{Filtering of connectivity matrices}
Connectivity matrices represent the degree of synchronization between all pairs of electrodes for one patient, and are analyzed separately for each frequency band. To reduce noise and improve data clarity, a sparse graph is generated from each synchrony matrix, following the method presented in Ref.~\cite{8_DeVicoFallani2017}: PLV matrices are filtered by applying a threshold to cancel a percentage of the weakest values, such that the final networks reached a predetermined mean degree, set here to 3, as recommended for small-size networks. Theoretical and numerical analyses indicate that this filtering method effectively preserves the core structure of weighted graphs, maintaining their hierarchical organization. It enables clear differentiation between groups and facilitates the tracking of rapid temporal changes in network organization~\cite{8_DeVicoFallani2017}. This method reduces the graph to a simplified and sparse form \( G = (\Omega, W) \) with \(n\) nodes (i.e number of electrodes implanted) with a weighted matrix \( W = \{w_{ij}\}_{i, j \in \Omega} \) that is symmetric and with $w_{ij} > 0$, denoting the PLV value between the pair of electrodes $(i,j)$.

\begin{figure*}[p]
    \centering
    \includegraphics[width=0.9\textwidth]{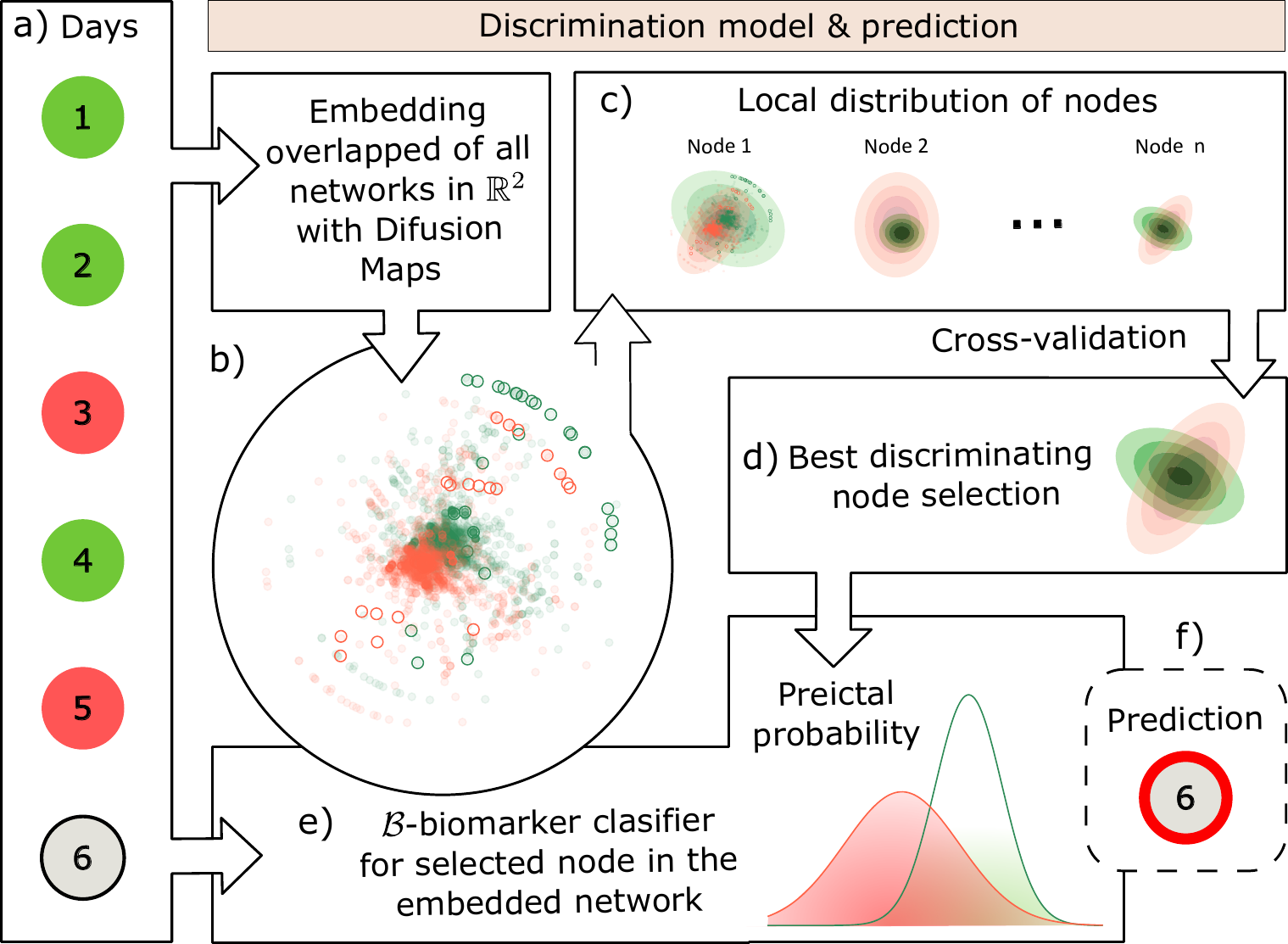}
    \caption{
    Illustration of ictal state prediction based on training from previous days, for a given patient and frequency band. The training process follows steps a)–b)–c)–d), and the testing process follows steps e)–f).    a) Segments from all days are labeled a priori and color-coded as green (interictal) or red (preictal). b) Each segment is embedded into a two-dimensional Euclidean space using diffusion maps, revealing its characteristic spatial organization. c) For each node, independent Gaussian distributions are estimated for the preictal and interictal states. d) Nodes that exhibit the highest discriminative power between both states are identified through cross-validation. e) In the testing phase, the unknown segment is embedded and classified as preictal or interictal based on the biomarker \(\mathcal{B}\). f) The ictal state of the entire day is determined by averaging the classification outcomes of all segments.
    }
    \label{fig:methodology}
\end{figure*}

\subsection{Euclidean embedding of networks}
This study employs an embedding framework within Euclidean space to represent the iEEG connectivity graphs from the preictal and interictal epochs. The Euclidean embedding was chosen to exploit the explicit metric structure inherent to this space, enabling precise quantitative analysis and leveraging the extensive repertoire of well-established statistical parametrization techniques. To achieve dimensionality reduction while preserving the intrinsic data geometry, we adopted the Diffusion Maps algorithm, a method that effectively captures the underlying manifold structure of high-dimensional data~\cite{coifman2006diffusion, lafon2006diffusion}.

For a connectivity graph $G$ the algorithm estimates the transition probability matrix $P$ of a Markov chain with entries $p_{ij}=\frac{w_{ij}}{d_{i}}$, where the strength of each node $d_i = \sum_{j \in \Omega} w_{ij}$.  Each entry $p_{ij}$ encodes thus the probability of moving from node $i$ to node $j$ through a random walk of length $1$. The random walk gives rise to a geometric diffusion with an associated distance between nodes $i$ and $j$ defined as~\cite{coifman2006diffusion, lafon2006diffusion}: $d^2_{ij} =\sum_{k\geq 0}\frac{p_{ik}-p_{jk}}{\mu_k^*}$, where the term $\mu_k^*$ denotes the unique stationary distribution of the Markov chain $P$. By construction, this diffusion distance between nodes is strongly ruled by the connectivity of the graph, and it takes small values if nodes are connected by many paths.

Considering the spectral representations of matrix $P$, one has a set of eigenvalues $|\lambda_0| \geq |\lambda_1| \geq \ldots \geq |\lambda_{N-1}|$ and eigenvectors $\varphi_k$ and $\psi_k$ such that $\varphi_k^TP = \lambda_k \varphi_k^T $ and $P\psi_k=\lambda_k \psi_k$. The diffusion distance can be written as: $d^2_{ij} = \sum_{k \geq 1}\lambda_k^{2}(\psi_k(i) - \psi_k(j))^2$ where $\psi_k(j)$ denotes the  component  $j$ of eigenvector $k$.

The diffusion distance can be approximated to a relative precision using the first $q$ nontrivial eigenvectors and eigenvalues ($\varphi_0 = \mu^*$ and $\psi_0$ is a constant vector): 
$d^2_{ij} \simeq \sum_{n = 1}^{q} \lambda_n^{2}(\psi_n(i) - \psi_n(j))^2 $. The diffusion map is then constructed as
\begin{equation}
    \Psi : x \mapsto 
    \begin{pmatrix}
    \lambda_1 \psi_1(x) \\
    \lambda_2 \psi_2(x) \\
    \vdots \\
    \lambda_q \psi_q(x)
    \end{pmatrix}.
\end{equation}

This mapping \( \Psi : \Omega \to \mathbb{R}^q \) is equivalent to projecting the set of nodes \( \Omega \) as a cloud of points in an Euclidean lower-dimensional space, where the rescaled eigenvectors and eigenvalues define the spatial coordinates~\cite{lafon2006diffusion}. This process effectively embeds the graph into $\mathbb{R}^q$ such that the diffusion distance between nodes is approximated by the Euclidean distance between their corresponding embedded representations.

An analysis of the eigenvalues derived from the transition matrices \(P\) revealed no distinct eigengap that would naturally suggest an optimal embedding dimension \(q\). However, our results show that, on average, the first two nontrivial eigenvectors capture over $80\%$ of the variance associated with the low-dimensional processes underlying brain dynamics. Based on this, we consistently selected \(q = 2\) throughout the study.
This choice enables a meaningful geometric representation of preictal and interictal states in \(\mathbb{R}^2\), guided primarily by the need for interpretability and visual inspection of the embedded graphs. The resulting low-dimensional embedding supports a geometric analysis of brain states, facilitates the construction and evaluation of our biomarker \(\mathcal{B}\), and provides a basis for comparison with more complex embedding techniques applied to the same dataset~\cite{guillemaud2024hyperbolic}. Beyond interpretability, the choice $q = 2$ is also motivated by statistical constraints inherent to the dataset. Since our analysis 
operates at the level of individual nodes, each Gaussian distribution 
is estimated from $n = 30$ embedded observations per day. Estimating a 
$q$-dimensional Gaussian requires $q + \frac{q(q+1)}{2}$ parameters 
(5 in $q=2$, 9 in $q=3$, 14 in $q=4$), and reliable covariance estimation 
demands $n \gg q$, with practical recommendations of 
$n \geq 10q$. With $n = 30$, this condition 
is only borderline satisfied for $q = 2$, and is no longer met for 
higher dimensions, where increasing $q$ under a fixed sample size leads 
to a systematic increase in covariance estimation error scaling 
approximately as $\mathcal{O}(\sqrt{q/n})$~\cite{vershynin2012close}. Consequently, while 
additional embedding components could in principle carry discriminative 
information, their exploitation would require substantially larger 
numbers of segments per day than are available in the present 
dataset.

Diffusion map-based embedding offers the advantage of providing a meaningful representation of the graph, while also defining an explicit distance in the space $\mathbb{R}^ q$ that reflects network’s connectivity strength. Applied to networks estimated from magnetoencephalographic (MEG) signals, diffusion map embedding allowed to highlight significant differences between the resting-state connectivity structure derived from epileptic patients and those observed in healthy controls~\cite{Chavez2010}. Additionally, in a study with a single patient, an extension of the methodology has been proposed to capture notable spatial shifts in the embeddings of long term iEEG signals, highlighting their potential for detecting changes prior to seizures~\cite{Duncan2013}.

\subsection{Alignment of embedded networks}
Prior to comparing groups of embedded networks, we aligned all the projected networks of each patient. This step is crucial, as structurally similar connectivity networks can yield substantially different embedding coordinates, thereby hindering the correspondence of networks across different days~\cite{richiardi2013machine}. Indeed, the eigenvalue decomposition used in the embedding method can result in coefficient sign flips, rotations, or variations in the ordering of the components. A minor connectivity perturbation may thus result in two embeddings with the same similarities between nodes but with a different structure in their coordinate representations~\cite{gursoy2023alignment}. Such effect may artificially inflate the distances between corresponding nodes in embeddings from different days, without reflecting genuine changes in network connectivity. To address this, an alignment procedure was applied to the brain network embeddings within  \(\mathbb{R}^2\), ensuring their comparability across samples.

In our study we employed the Procrustes technique as this method modifies the data through translations, rotations, reflections, and scaling, while preserving its main geometric organization~\cite{gower1975generalized}. Specifically, we use Generalized Procrustes Analysis (GPA), which belongs to a family of methods designed for analyzing multivariate data to obtain a consensual configuration among several datapoints (here, the embedded networks)~\cite{dijksterhuis1991interpretation}. Briefly,  GPA transforms a matrix \(\mathbf{A}\) of dimension \(m \times n\) to be as close as possible to another matrix \(\mathbf{B}\) of the same dimensions through the element \(\mathbf{AT}\), where \(\mathbf{T}\) is an \(n \times n\) matrix that minimizes the trace \(
\operatorname{tr}[(\mathbf{A} \mathbf{T} - \mathbf{B})'(\mathbf{A} \mathbf{T} - \mathbf{B})]\). This optimization is achieved through the singular value decomposition (SVD) of the main matrices~\cite{cuadras2007nuevos}. The matrix element \(\mathbf{T}\) is then defined as $\mathbf{T} = \mathbf{U} \mathbf{V}'$. 

\begin{figure}[!htbp]
    \centering
    \includegraphics[width=0.9\textwidth]{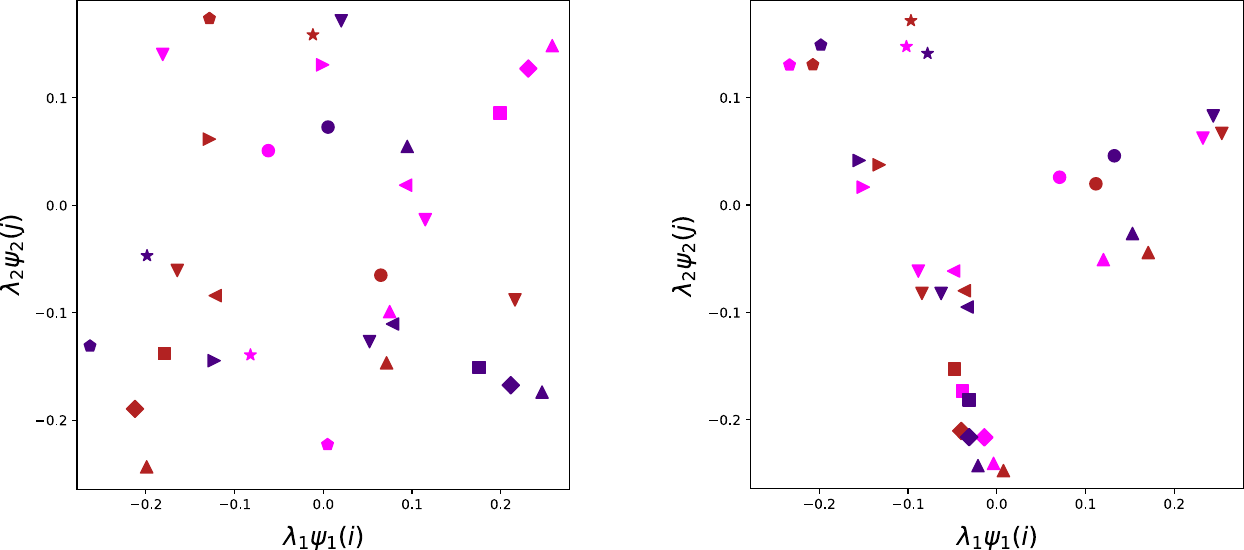}
    \caption{Alignment of embedded networks. (Left) Euclidean embedding of three consecutive network segments for patient 5: segment t (red), t+1 (magenta), and t+2 (purple). Each point represents a node (electrode) in the embedding space, with different marker shapes used to distinguish individual nodes, such that same markers correspond to the same node across different time segments. (Right) Procrustes alignment of the corresponding preictal embeddings, illustrating the spatial consistency after alignment, as markers corresponding to the same node across different time segments tend to cluster together.}    
    \label{fig:Embed_Procrus}
\end{figure}

A set of matrices \(\mathbf{X}\) of dimension \(n \times 2\) represents \(n\) pairs of embedding coordinates in the \(\mathbb{R}^2\) of a network from a single patient. These are transformed to approximate a matrix element \(\mathbf{Y}\) of the same dimension, representing a single reference segment of the same patient selected from the set of interictal segments. In our specific case, for each interictal segment, Procrustes alignment of all other embeddings (both interictal and preictal) is performed against it and compute the mean squared error (averaged over all the aligned segments). The segment that minimizes  this total alignment error is retained as the reference. This procedure 
is applied independently for each patient, band, and evaluation day, 
respecting the data availability of each evaluation scheme: in the 
leave-one-out classification, the reference is computed from all days 
except the one being evaluated, whereas in the pseudo-prospective 
prediction, only data from preceding days are used. The transformation is established as $\mathbf{Y}^* = b \mathbf{X} \mathbf{T} + \mathbf{C}$. Here, \(\mathbf{Y}^*\) is the transformed matrix that minimizes the pointwise distances between the matrices \(\mathbf{X}\) and \(\mathbf{Y}\). The scalar \(b\), which accounts for distortion-free scaling of the datasets, is set to \(b = 1\). \(\mathbf{C}\) is a translation vector determined by the barycenters \(\overline{\mathbf{x}}\) and \(\overline{\mathbf{y}}\). This non-scaled transformation of a network's nodes is illustrated in \cref{fig:Embed_Procrus}.

\section{Models for embedded nodes}
In this study, to evaluate the potential of preictal states discrimination within the Euclidean space \(\mathbb{R}^2\) resulting from the diffusion map embedding of brain networks, we adopt a local (node-based) perspective. We aim to identify nodes (i.e. electrodes) whose connectivity highlights them as key nodes of interest for preictal state identification.

To achieve this, we performed an aligned embedding in \(\mathbb{R}^2\) of all networks across all interictal and preictal epochs for each patient, effectively creating a superposition of \(30 \times d \times n\) points projections, where \(d\) represents the number of recorded days, each consisting of 30 connectivity networks, where each embedded graph has \(n\) points corresponding to \(n\) intracranial electrodes. Since a previous study has suggested that the spatial distribution of embedded nodes significantly varies prior to seizure~\cite{duncan2013intracranial}, we proceed to locally parameterize and quantify these distributional differences. To this end, we extracted the subset of \(30 \times d\) projections for each node to facilitate a comparative analysis between interictal and preictal networks, as illustrated in~\cref{fig:methodology}. 

Here, we aimed at detecting nodes from the preictal states as outliers that differ from their distribution observed during interictal baseline. The Bhattacharyya distance was used as a parameter to identify outlier nodes for each patient. This is an important measure for quantifying the separability between two normal distributions \cite{fukunaga2013introduction}. Since it is a metric in Euclidean space, its interpretation is straightforward: the greater the Bhattacharyya distance, the larger the difference in the spatial nodes' distribution from the nodes' organization during seizure-free epochs. This metric is defined as:
\begin{equation}
\begin{aligned}
D_b = & \frac{1}{8} (\boldsymbol{\mu}_2 - \boldsymbol{\mu}_1)^T 
\left[ \frac{\boldsymbol{\Sigma}_1 + \boldsymbol{\Sigma}_2}{2} \right]^{-1} 
(\boldsymbol{\mu}_2 - \boldsymbol{\mu}_1) \\
& + \frac{1}{2} \ln \frac{\left| \frac{\boldsymbol{\Sigma}_1 + \boldsymbol{\Sigma}_2}{2} \right|}
{|\boldsymbol{\Sigma}_1|^{1/2} |\boldsymbol{\Sigma}_2|^{1/2}},
\end{aligned}
\end{equation}
where \(\boldsymbol{\mu}_i\) and \(\boldsymbol{\Sigma}_i\) correspond to the mean vectors and covariance matrices of the preictal and interictal distributions of a given node. 

\subsection{Impact of the alignment}
To assess the statistical significance of the Bhattacharyya distance $D_b$ for each node, a permutation test was conducted on the already-aligned embeddings. For each patient, band, and evaluation day, interictal and preictal segment labels were randomly permuted, and $D_b$ was recomputed per node, testing the null hypothesis that the two states share the same local spatial distribution in the embedding space. Permutation-based $p$-values were estimated for each node, and only those with $p < 0.05$ (FDR-corrected) were retained as statistically significant. 

\subsection{Preictal biomarker \(\mathcal{B}\)}
This study proposes the implementation of the parameter \(\mathcal{B}\) as the final step for preictal state discrimination, based on the nodal distributions learned from previous interictal and preictal states in each epilepsy patient (\cref{fig:methodology}). To achieve this, a connectivity network associated with an unknown state is embedded and aligned in the Euclidean space. The points corresponding to the most discriminant nodes (previously identified) are compared against the reference nodal distributions of both ictal states. The classification is performed using the evaluation of the multivariate probability density functions (PDF) of each point under each of the two distributions. The PDF of a node \( \mathbf{x}_j \) under the distribution \(s \in \{\text{interictal}, text{preictal}\}\) is given by:

\begin{equation}
    PDF_s(\mathbf{x}_j) = \frac{1}{(2\pi) 
    |\boldsymbol{\Sigma}_i^s|^{1/2}} 
    \exp \left( -\frac{1}{2} (\mathbf{x}_j - \boldsymbol{\mu}_i^s)^T 
    (\boldsymbol{\Sigma}_i^s)^{-1} (\mathbf{x}_j - 
    \boldsymbol{\mu}_i^s) \right)
\end{equation}

Based on this, we define the proposed biomarker \(\mathcal{B}\) as a dimensionless parameter that classifies the state of a given network node \(\mathbf{x}\), based on the likelihoods under the preictal and interictal reference distributions.

\[
\mathcal{B} = \frac{\text{PDF}_{\text{inter}}(\mathbf{x})}
{\text{PDF}_{\text{pre}}(\mathbf{x})}
\]
with the classification criterion of state \(s\) given by:
\[
s(\mathbf{x}) =
\begin{cases}
\text{preictal}, & \text{if } \mathcal{B}(\mathbf{x}) \leq 1, \\
\text{interictal}, & \text{if } \mathcal{B}(\mathbf{x}) > 1.
\end{cases}
\]

The threshold $\mathcal{B} \leq 1$ corresponds to a likelihood ratio test under equal prior probabilities. This choice is a principled consequence of the pseudo-prospective forecasting framework adopted in this study, where the classifier operates exclusively on information available at the time of prediction.

\subsection{Discrimination between interictal and preictal states}
The classification of a given network is evaluated using the Leave-One-Out Cross-Validation prediction method, where all segments of the day to be tested are excluded from the training phase. The full pipeline is subsequently applied to the held-out day, including the embedding of the corresponding connectivity matrices and their alignment with respect to the reference segment.

\begin{figure*}[!htbp]
    \centering
    \includegraphics[width=1\textwidth]{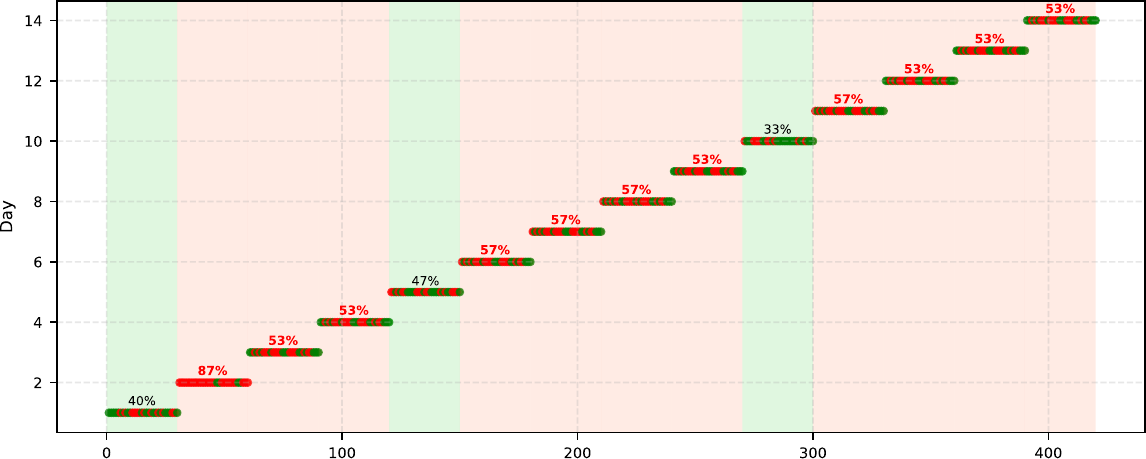}
    \caption{Leave-one-out cross-validation prediction for patient 6 in the Low \(\gamma\) band. A priori known ictal states are indicated by a colored background, green: interictal, red: preictal. Each day displays the prediction for its 30 networks obtained from the PLV connectivity matrices. The numerical values represent the percentage of segments predicted as preictal. We classify these percentages as preictal state if \(\geq 50 \%\), and seizure-free otherwise.}
    \label{fig:prediction from classification}
\end{figure*}

Although nodes surviving the FDR-corrected threshold ($p < 0.05$) may be considered statistically significant candidates for inclusion in the model, statistical significance alone does not determine their selection. Rather, node selection remains driven by each node's contribution to classification and predictive performance. To further control for false positives, only $\lceil N \times 0.30 \rceil$ significant nodes were retained, where $\lceil \cdot \rceil$ denotes the ceiling function and $N$ is the total number of significant nodes. This proportion was determined empirically by evaluating the effect of varying the percentage of retained significant nodes on overall classification performance. The mean of the $\mathcal{B}$ values across the selected nodes is subsequently computed to determine the predicted class. Each segment is classified as belonging to either the preictal or interictal class according to the biomarker criterion, and this procedure is repeated iteratively across all days for each patient. The classification probability for each day is computed as the proportion of segments labeled as preictal, as illustrated in \cref{fig:prediction from classification}. 

A decision threshold of $50\%$ was applied to assign a day to the preictal class, such that a day was classified as preictal if at least half of its constituent segments were labeled accordingly. This threshold implicitly assumes equal prior probabilities for the two classes and constitutes a principled choice within the pseudo-prospective forecasting framework adopted in the present study, in which the classifier operates exclusively on information available at the time of prediction, thereby precluding any form of look-ahead bias.

The model's effectiveness in distinguishing preictal from interictal states was assessed using standard classification metrics: the F1-score, and the balanced accuracy. Given the class imbalance in the dataset, where the number of interictal and preictal days is not evenly distributed, the F1-score was selected as a key evaluation metric, as it balances the trade-off between false positives and false negatives.  The balanced accuracy was also calculated, which is defined as the average of sensitivity and specificity, which ensured a fair evaluation of the model's performance across both classes, also mitigating the impact of the dataset's imbalance. 

The most discriminative frequency band for each patient was selected based on its classification performance metrics, independently evaluated for each band. Notably, the bands containing nodes with the lowest $p$-values do not necessarily yield the highest discriminative performance; therefore, the individual patient-level classification metrics per band serve as the primary criterion for band selection. For the most discriminative band for each patient, a group-level permutation test was conducted to assess whether the observed classification performance exceeded chance level. To this end, day labels were shuffled independently within each patient, and performance metrics were subsequently averaged across patients over their respective most discriminative bands.

\subsection{Daily seizure risk forecasts}
Finally, we performed a pseudo-prospective analysis for epilepsy episode prediction. In this forecasting approach, consecutive days were included in the training phase until at least one day of each class was present. During the training phase, statistically significant nodes were first identified through the permutation-based procedure, and the $\lceil N \times 0.30 \rceil$ most significant nodes were subsequently retained based on their individual contributions to classification performance. The biomarker $\mathcal{B}$ was then computed for each segment of the immediately following day in order to perform classification. The predicted class probability of the day was determined by the proportion of segments classified as preictal among the 30 segments of the predicted day. A threshold of \( \geq 50\%\) was used to classify the day as preictal, while lower values were assigned to the interictal state.

To provide a full evaluation of forecasting models, we computed the accuracy and Brier score. The accuracy provides a straightforward metric to assess the overall correctness of daily predictions in a range of $[0,1]$ being $1$ the most accurate, while the Brier score  (i.e., the mean squared error over every forecast) offers a complementary view by quantifying the probabilistic calibration of the model and the confidence of the classifier in its predictions on a range $[0,1]$ being $0$ the most confident. 

\section{Results}
\begin{figure}[!htb]
    \centering
    \includegraphics[width=0.7\textwidth]{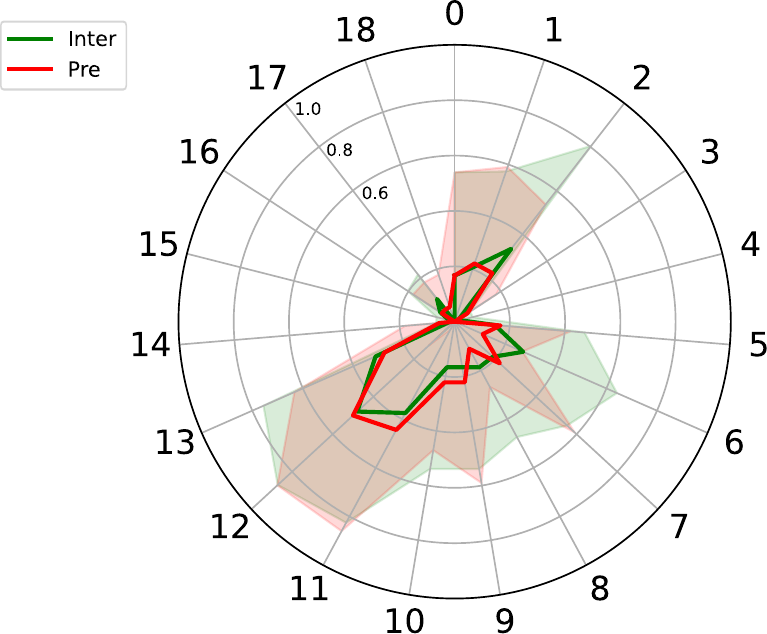}
\caption{Mean prevalence of each node as statistically significant ($p < 0.05$, FDR-corrected) across all frequency bands for patient 2 (from 0 to 18 nodes/electrodes). Thick lines represent the mean prevalence, and shaded regions indicate the corresponding standard deviation across cross-validation runs, for the interictal (green, 8 days) and preictal (red, 3 days) states.
} \label{fig: histograma_reloj}
\end{figure}

The proposed methodology focuses on the localization of discriminating nodes in epilepsy patients through an Euclidean space embedding, aiming to differentiate between interictal and preictal brain connectivity. Through the calculation of the Bhattacharyya distance between nodal projections and evaluating its statistical relevance through permutation-based $p$-values, it is revealed that a subset of electrodes/nodes consistently exhibits significant geometric reorganization in the Euclidean space. A key finding is that the identity of the most discriminant nodes remains highly consistent across cross-validation runs, as illustrated in \cref{fig: histograma_reloj}. In other words, the significant nodes remain consistent across days for a given patient across states in the different frequency bands..

\begin{table}[h]
    \centering
    \caption{Discrimination between interictal and preictal resting states: F1 score and balanced accuracy. Global average across all bands and patients, and most discriminative band per patient.}
    \begin{tabular}{c c c}
        \hline
        \hline
        \textbf{} & \textbf{F1 score} & 
        \textbf{Balanced accuracy} \\
        \hline
        Average across all bands & 
        0.45 $\pm$ 0.36 & 
        0.54 $\pm$ 0.23 \\
        Most discriminative band & 
        0.64 $\pm$ 0.32 & 
        0.78 $\pm$ 0.16 \\
        \hline
        \hline
    \end{tabular}
    \label{tab: band discrimination}
\end{table}

\subsection{Discrimination of preictal states}
When classification performance is averaged across all frequency bands and all patients, the F1 score and the balanced accuracy, remain close to random level classification, as shown in~\cref{tab: band discrimination}. These overall classification performances indicate that no single frequency band is universally optimal, a finding consistent with the known heterogeneity of epileptic networks. However, closer inspection of each patient reveals distinct frequency bands where the model is highly discriminant. 

It becomes clear that some frequency bands allow a better classification on an individual level. For example, as shown in~\cref{fig:prediction from classification}, the model achieved a high discrimination of seizure and seizure-free days for patient~6 in the Low~$\gamma$ band. To highlight this effect, the last row in~\cref{tab: band discrimination} shows the average classification metrics computed only on the most discriminative frequency band for each patient. 
This selection yields a notable improvement in classification performance, with an F1-score of $0.64 \pm 0.32$ and a balanced accuracy of $0.78 \pm 0.19$, suggesting that the proposed method is capable of capturing meaningful spectral information for seizure state classification when the most discriminative frequency band is considered. The group-level permutation test confirms that these results are significantly above chance level for both balanced accuracy and F1-score ($p = 0.001$).

To contextualize these results, \cref{tab: comparison} presents a comparison of our approach against two established methods evaluated on the same dataset, the same patient cohort, and the same set of implanted electrodes within the clinically identified seizure onset zone (SOZ): (i) an SVM classifier applied to PLV connectivity values~\cite{cousyn2023daily}, and (ii) a hyperbolic embedding-based classifier~\cite{guillemaud2024hyperbolic}. All methods thus operate under identical informational constraints.

\begin{table}[h]
    \centering
    \caption{Comparative performance of classification metrics for different methodological approaches applied to the same dataset. Missing values were not reported in the original papers.}
    \begin{tabular}{l c c c c}
        \cline{2-4}
        & \multicolumn{3}{c}{\textbf{Classification}}  \\
        \cline{2-4} 
        & \textbf{F1-Score} & \textbf{Bal.\ Acc.} & 
        \textbf{Accuracy}  \\
        \hline
        Euclidean & $0.64 \pm 0.32$ & $0.78 \pm 0.16$ & $0.78 \pm 0.19$ \\
        Hyperbolic & $0.78 \pm 0.18$ & -- & $0.77 \pm 0.13$ \\
        SVM & $0.79 \pm 0.09$ & -- & $0.85 \pm 0.09$ \\
        \hline
    \end{tabular}
    \label{tab: comparison}
\end{table}

\subsection{Forecasting of preictal days}
To assess the real-world predictive utility of the proposed biomarker $\mathcal{B}$, a pseudo-prospective forecasting analysis was conducted following the procedure outlined in~\cref{fig:methodology}. In this framework, the model predicts the ictal state of a given day using only data from preceding days, ensuring that the training set includes at least one day from each ictal class.  The global forecasting performance averaged across all patients and the performance for the most discriminatory band per patient is shown in~\cref{tab: band discrimination forecasting} . Similar to the results observed in the classification task, the forecasting results reflect the fact that no single frequency band is universally optimal. 

\begin{table}[h]
    \centering
    \caption{Pseudo-prospective forecasting of interictal and preictal states: Accuracy and Brier score. Global average across all bands and patients and most discriminative band per patient.}
    \begin{tabular}{c c c}
        \hline
        \hline
        \textbf{} & \textbf{Accuracy} & \textbf{Brier score} \\
        \hline
        Average across  all bands & 
        0.49 $\pm$ 0.33 & 
        0.28 $\pm$ 0.14 \\
        Most discriminative band & 
        0.86 $\pm$ 0.18 & 
        0.13 $\pm$ 0.11 \\
        \hline
        \hline
    \end{tabular}
    \label{tab: band discrimination forecasting}
\end{table}

\begin{figure*}[ht]
    \centering
    \includegraphics[width=1\textwidth]{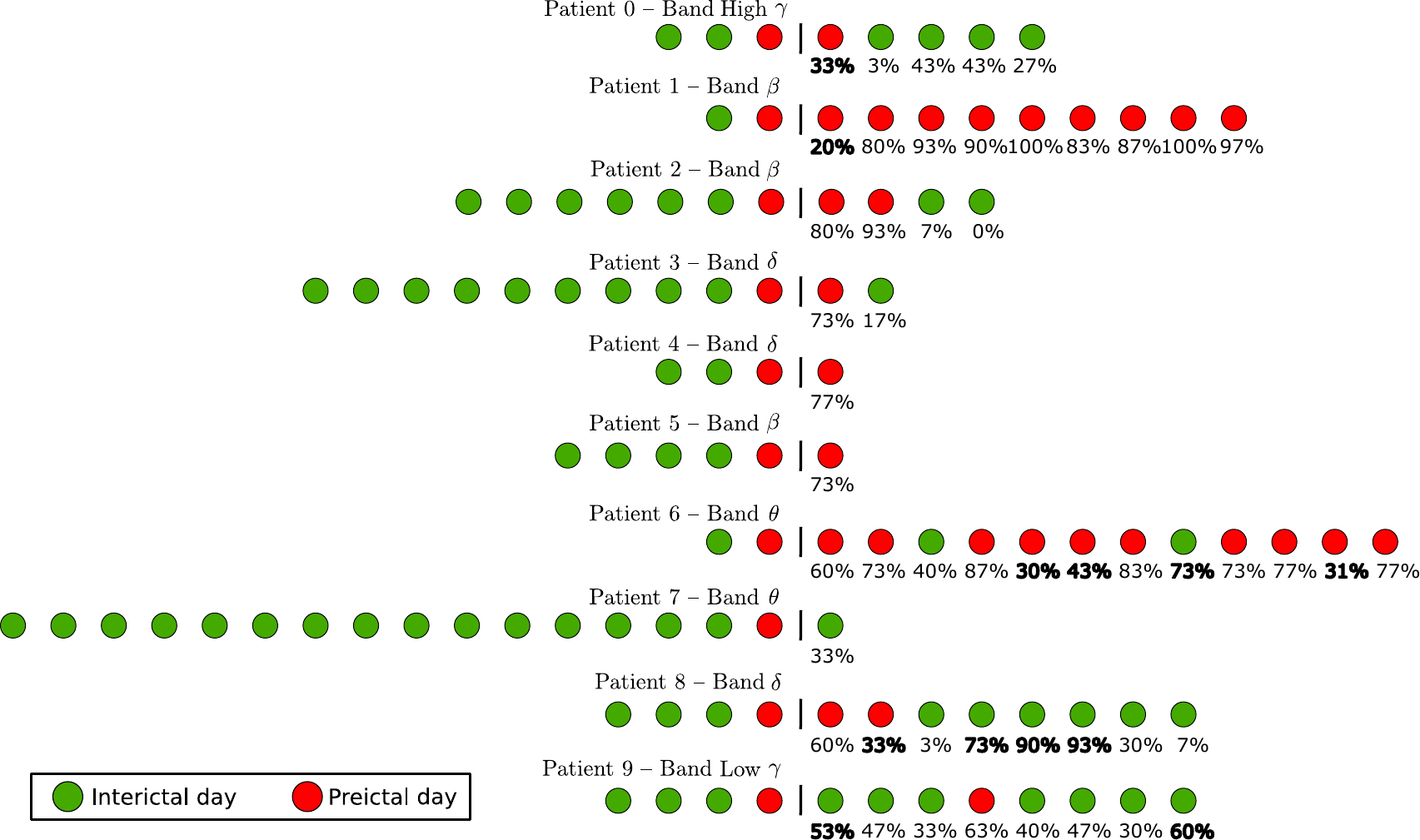}
    \caption{Forecasting performance in the pseudo-prospective 
    prediction application, where all previous days were used as 
    training to predict the next day. The best performance for each 
    patient is presented with their most discriminative frequency band. 
    Probabilities that failed to predict the correct class are shown 
    in bold.}
    \label{fig: all_predictions}
\end{figure*}

However, the patient-specific forecasting results shown in~\cref{fig: all_predictions} reveal a different picture, demonstrating strong predictive capabilities when the model is applied to the most informative frequency bands.  When the evaluation is restricted to these patient-specific high-performance bands, the classification metrics improve substantially, reaching an average accuracy of $0.86 \pm 0.18$ and a low Brier score of $0.13 \pm 0.11$. A group-level permutation test yielded $p = 0.026$ for accuracy and $p = 0.029$ for the Brier score, confirming that the overall forecasting performance is significantly better than chance level.

\cref{tab: comparison prediction} presents the prediction performance comparison against the same methods mentioned before. Results show that the proposed Euclidean embedding method achieves prediction performance competitive with both the hyperbolic embedding and the SVM.

\begin{table}[h]
    \centering
    \caption{Comparative performance of prediction metrics for 
    different methodological approaches applied to the same dataset. 
    Missing values were not reported in the original 
    papers.}
    \begin{tabular}{l c c}
        \cline{2-3}
        & \multicolumn{2}{c}{\textbf{Prediction}} \\
        \cline{2-3}
        & \textbf{Brier Score} & \textbf{Accuracy} \\
        \hline
        Euclidean & $0.13 \pm 0.11$ & $0.86 \pm 0.18$ \\
        Hyperbolic & $0.12 \pm 0.12$ & $0.87 \pm 0.17$ \\
        SVM & $0.13 \pm 0.09$ & -- \\
        \hline
    \end{tabular}
    \label{tab: comparison prediction}
\end{table}

From the total of 28 days labeled as preictal, 22 were correctly classified by the proposed forecasting model, yielding a true positive rate (TPR) or sensitivity of $78.57\%$. For the 23 interictal days, 17 were accurately identified, resulting in a specificity of $73.91\%$. On the other hand, 6 interictal days were misclassified as preictal (false positives), and 6 preictal days were misclassified as interictal (false negatives). These metrics provide a strong indication that the proposed methodology shows good discriminative capacity of a patient-specific model to anticipate early detection of preictal states, which is the primary goal in seizure forecasting.

\section{Discussion}
This study demonstrates that low-dimensional representations of iEEG connectivity networks, derived from daily vigilance-controlled resting-state recordings, can effectively distinguish between seizure and seizure-free days, offering a reliable and interpretable tool for forecasting daily seizure risk. The use of vigilance-controlled recordings minimizes fluctuations in brain connectivity due to varying alertness levels, thereby enhancing the robustness of preictal state detection. Unlike conventional seizure prediction approaches that rely on continuous long-term EEG monitoring, our results show that brief, daily resting-state sessions are sufficient to capture preictal dynamics at the daily scale~\cite{cousyn2023daily, guillemaud2024hyperbolic}. By embedding brain connectivity into a low-dimensional Euclidean space, the proposed methodology further enables the identification of specific network nodes that differentiate interictal from preictal states. This feature holds considerable clinical relevance, as it supports the development of individualized monitoring and intervention strategies aimed at anticipating seizures. Overall, these findings underscore the promise of low-dimensional network embeddings as practical biomarkers for real-time, patient-specific seizure risk assessment.

The proposed model advances seizure forecasting based on brain connectivity analysis within a low-dimensional vector space. While more complex embedding approaches---such as those based in hyperbolic spaces---may offer enhanced predictive performance~\cite{guillemaud2024hyperbolic}, our results support the effectiveness of Euclidean embeddings in capturing dynamic changes associated with epilepsy. As reported in~\cref{tab: comparison,tab: comparison prediction}, the proposed Euclidean embedding achieves prediction performance comparable to that of both the hyperbolic embedding approach and the SVM-based classifier, with all methods evaluated on the same dataset and under identical experimental and informational constraints. While the original dataset did not reveal any definitive links between seizure activity and medication tapering~\cite{cousyn2023daily}, the potential impact of timing and treatment modifications cannot be entirely dismissed---an inherent challenge in clinical research. These factors may act as confounding variables and should be carefully considered when interpreting the findings.

Given the well-defined metric structure of Euclidean space, our approach facilitates a more interpretable analysis of network dynamics and seizure-related connectivity changes. Moreover, our results support previous works suggesting that embedding brain connectivity networks into lower-dimensional spaces is sufficient for forecasting seizure risk~\cite{Duncan2013, guillemaud2024hyperbolic}. However, further research is needed to determine the specific conditions under which other geometrical embeddings may offer advantages, particularly in cases where hierarchical organization plays a dominant role in network connectivity.

Our findings indicate that the most discriminating nodes remain consistent across different days, suggesting that these nodes may play a fundamental role in the neural mechanisms underlying seizure susceptibility. This observation is in line with recent work showing that ordinal complexity metrics can identify spatially localized electrode-level signatures of seizure imminence~\cite{granado2025canine}, further supporting the relevance of node-level characterization for preictal state detection.

As demonstrated in the classification and forecasting analyses, the most discriminative band for each patient was selected based on performance metrics independently evaluated per band, and could be identified after an initial calibration period in a prospective setting. The model's sensitivity to frequency bands highlights the individualized nature of seizure dynamics where most of the useful discriminant information seems to be contained in the connectivity at the $\delta$ and $\alpha$ bands. The \(\delta\) and $\beta$ bands exhibited the highest prediction performance across all patients, which aligns with prior studies emphasizing the role of such oscillations in seizure-related neural activity~\cite{cousyn2023daily}. These findings are consistent with recent information-theoretic approaches showing that frequency-specific biomarkers, derived from high-frequency oscillations~\cite{granado2022high} and multiscale symbolic analysis of delta oscillations~\cite{granado2024multiscale}, carry significant discriminative power for preictal state characterisation. Nevertheless, our results suggest that optimal frequency band selection should be tailored to individual patients rather than assuming a universal optimal band.

The pseudo-prospective forecasting approach demonstrated high prediction sensitivity ($78.57\%$), yet exhibited a tendency to misclassify preictal days. Furthermore, false positives were observed in $26\%$ of interictal days, indicating that some interictal segments may share connectivity properties with preictal states. This raises the question of whether these misclassifications represent a genuine high susceptibility of seizures.

Beyond classification and forecasting, this methodology contributes to a broader understanding of how brain connectivity changes prior to seizure dynamics. The ability to track network reconfigurations within a low-dimensional space provides new insights into the spatiotemporal evolution of seizure-related activity. Additionally, the patient-specific identification of discriminative nodes suggests the potential for personalized seizure prediction models. Furthermore, characterizing seizure-related changes in connectivity at the level of individual nodes may provide valuable biomarkers for assessing treatment responses in patients undergoing epilepsy surgery or neuro-stimulation therapy. Nevertheless, given the heterogeneity of electrode location among the patients, we cannot generalize a single unique prediction model. We acknowledge that the cohort size (10 patients) and its heterogeneity in recording days and electrode count constitute a limitation of this study, inherent to the clinical constraints of intracranial EEG monitoring. Individual-level metrics for patients with very few evaluation days should be interpreted with caution. Studies involving significantly larger and more diverse patient cohorts are therefore needed to map out the limits and potential of our seizure risk forecasting.

Overall, this study advances the application of low-dimensional Euclidean embeddings and introduces a biomarker that serves as a promising tool for seizure classification and forecasting. Applied to multi-modal recordings, our approach could detect a daily and reliable signature of upcoming seizure(s), offering a broader time window for potential intervention or monitoring throughout the day.

\section*{Ethics}
Patients gave written informed consent (project C11- 16 and C19-55, conducted by INSERM and approved by the local ethic committees, CPP Paris VI and Sud Méditerranée 1).

\section*{Data Accessibility}
The network data that support the findings of this study are available on request from the corresponding author. The raw data are not publicly available due to privacy and ethical restrictions.

\section*{Authors’ Contributions}
S.R-A.: investigation, methodology, results visualization, writing—original draft, and editing; 
M.G.: data curation, investigation, methodology, writing—original draft;
A.L.: methodology, writing—original draft;
V.N.: data curation, writing—review and editing;
L.C.: data curation, methodology, writing—review and editing;
M.C.: conceptualization, supervision, result visualization, writing-original draft and editing;
All authors gave final approval for publication and agreed to be held accountable for the work performed therein.

\section*{Funding}
This study was supported by the program ``Investissements d'Avenir'' ANR-10-IAIHU-06, and grants from the ``Fondation de l'APHP pour la Recherche - Project PRIAM (Kniazeff Fund)''.

\section*{Conflict of interests}
V. Navarro reports fees from Boards with UCB Pharma, 408 EISAI, Liva Nova, GW Pharma.

\section*{Acknowledgements. }
S.R-A. gratefully thanks Luis A. Nu\~{n}ez for the guidance and mentorship throughout  the course of this work, and acknowledges financial support from the ERASMUS+ project, Latin-American alliance for Capacity buildiNG in Advance physics (LA-CoNGA physics). M.G. acknowledges doctoral support from the Ecole Normale Sup\'erieure de Lyon and Inria - Paris, France. A.L. acknowledges financial support from the doctoral school EDITE from Sorbonne University and Inria - Paris. 

\bibliographystyle{elsarticle-num} 
\bibliography{seizureForecastingEmbeddingREFS}
\end{document}